\documentclass[runningheads]{llncs}
\usepackage{graphicx}
\usepackage{subfig}
\begin{document}
\title{Estimating MRI Image Quality via Image Reconstruction Uncertainty}%\thanks{Supported by organization x.}}
\titlerunning{Image Reconstruction Uncertainty}
\author{Richard Shaw\inst{1,2} \and
Carole H. Sudre\inst{2,3} \and
Sebastien Ourselin\inst{3}\and
M. Jorge Cardoso\inst{3}}
%\author{***\inst{1} \and
%***\inst{2,3} \and
%***\inst{3} \and
%***\inst{3}}
%
\authorrunning{R. Shaw et al.}
%\authorrunning{***}
\institute{Dept. Medical Physics \& Biomedical Engineering, University College London, UK
\and
School of Biomedical Engineering \& Imaging Sciences, King’s College London, UK
\and
Dementia Research Centre, Institute of Neurology, University College London, UK}

%\institute{***
%\and
%***
%\and
%***}
%%
\maketitle              % typeset the header of the contribution
\begin{abstract}
Quality control (QC) in medical image analysis is time consuming and laborious, leading to increased interest in automated methods. However, what is deemed suitable quality for algorithmic processing may be different from human-perceived measures of visual quality. In this work, we pose MR image quality assessment from an image reconstruction perspective. We train Bayesian CNNs using a heteroscedastic uncertainty model to recover clean images from noisy data, providing measures of uncertainty over the predictions. This framework enables us to divide data corruption into learnable and non-learnable components and leads us to interpret the predictive uncertainty as an estimation of the achievable recovery of an image. Thus, we argue that quality control for visual assessment cannot be equated to quality control for algorithmic processing. We validate this statement in a multi-task experiment combining artefact recovery with uncertainty prediction and grey matter segmentation. Recognising this distinction between visual and algorithmic quality has the impact that, depending on the downstream task, less data can be excluded based on ``visual quality" reasons alone.

%\keywords{MRI \and quality control \and deep learning \and uncertainty}
\end{abstract}
\section{Introduction}

Most tasks in medical image analysis, and specifically deep learning, require a certain level of quality control (QC) of magnetic resonance imaging (MRI) data. QC is the process of establishing whether a scan or dataset meets a required set of standards, but it is usually based purely on the human-perceived ``visual quality’’ of the image rather than what is the acceptable level of image quality required for a particular algorithmic task, such as segmentation. We refer to this as the ``algorithmic quality” of the data. Indeed, what may be deemed of acceptable quality for a radiological assessment may not be sufficient to provide reliable measurements for some of the automated analyses the image would undergo, and vice versa. 

In the context of QC, MRI datasets can be affected by varied issues, most commonly acquisition noise, resolution, bias field, FOV/aliasing/wrap-around artefacts and motion artefacts. Other less common artefacts include zipper or radio-frequency (RF) spikes, blood-flow and metal artefacts. To date, the gold standard to identify such diversity of artefacts and ultimately decide whether to use or discard an image for downstream analyses remains the labour intensive visual inspection by experts \cite{Graham2018}. However, with the current trend of acquiring and exploiting very large imaging datasets, the time and resources required to perform this visual QC have become prohibitive. Furthermore, as with other rating tasks, visual QC is subject to inter and intra-rater variability due to differences in radiological training, rater competence, and sample appearance \cite{Sudre2019}. Some artefacts, such as those caused by motion, can also be difficult to detect, as their identification requires the careful examination of every slice in a volume. These challenges have led to an increased interest in automated methods. 

In this work, we propose to estimate uncertainty using a Bayesian deep learning framework for the auxiliary task of image reconstruction and show this can be used as a measure of ``algorithmic quality." Separating the learnable and non-learnable part of artefact correction for quality control we argue there is a distinction between QC for visual assessment and QC for algorithmic processing; apparent image corruption does not translate directly in the conclusion that the task cannot be performed. In consequence, recognising this distinction can ultimately result in a decrease in the number of images excluded for quality reasons.

\begin{figure}[!tbp]
  \centering
  \subfloat[]{\includegraphics[width=0.25\textwidth]{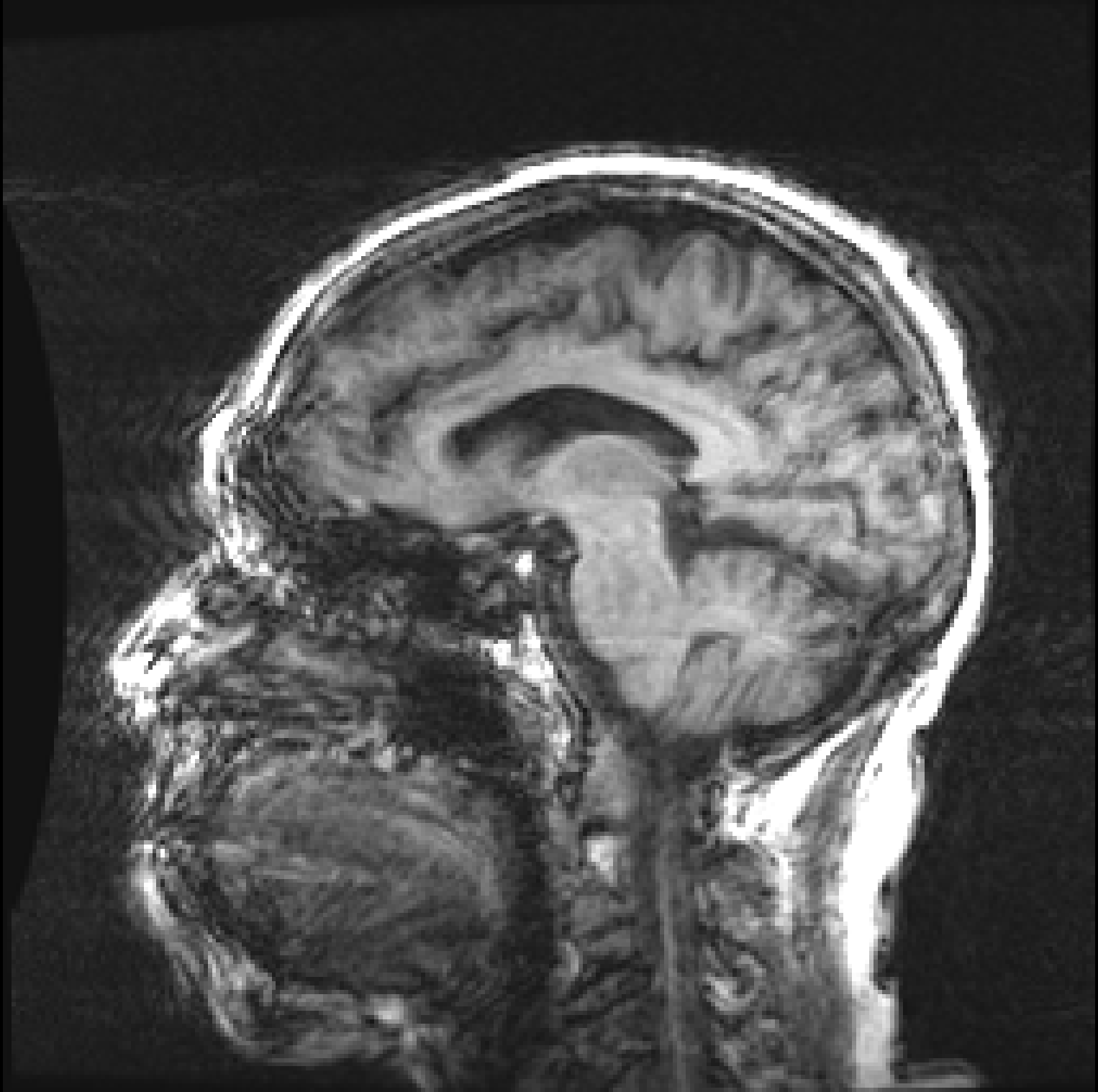}
  \label{fig:f1}}
  \hspace{1cm}
  \subfloat[]{\includegraphics[width=0.25\textwidth]{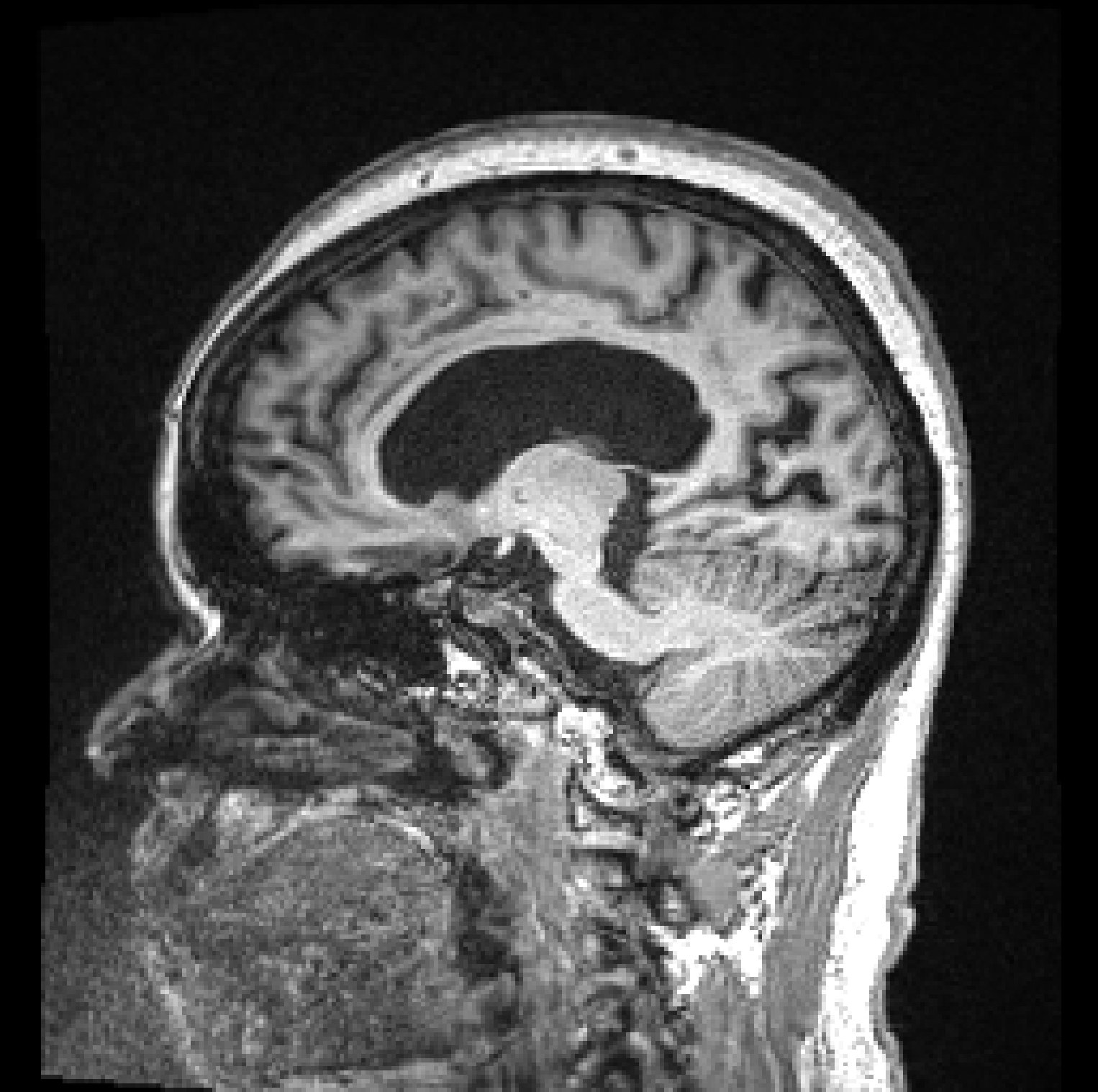}
  \label{fig:f2}}
  \caption{Two scans failing visual QC due to motion artefacts. a) The artefact located in the brain and affecting downstream gray matter segmentation task. b) Artefact localised at the bottom of the image, leaving the brain unaffected. From an algorithmic point of view, this image could have been used, highlighting the distinction between visual and algorithmic QC.}
  \label{fig:qc}
\end{figure}

 Fig. \ref{fig:qc} presenting two T1-weighted images from the ADNI database illustrates this difference. Both images have been graded by trained analysts and have been identified as poor quality and deemed ``unusable" (the criteria used for the QC is described in detail on the ADNI website \url{http://adni.loni.usc.edu/methods/mri-tool/mri-analysis/}) for ``containing artefacts due to motion." However, since on the right image the artefacts do not intersect with the brain, segmenting gray matter would not be affected and the scan should have been kept for this analysis. Thus, there is a clear distinction between what a human defines as acceptable scan quality and the quality required for further algorithmic processing. Furthermore, by training neural networks with noise/artefacts, CNNs make the internal representation of the data more robust to the presence of noise.  Therefore a certain level of image quality degradation may be tolerated before observing a drop in performance. In this work, we present a Bayesian neural network framework that allows us to measure image quality through the estimation of uncertainty associated with the task of image reconstruction.  After showing that the non-recoverable part of the corruption is associated with the level of uncertainty, we show how such representation is used in a multi-task setting and can inform the uncertainty of the associated task.

% \subsection{Objectives}

% Separating learnable and non-learnable part of artefactual corruption for quality control with the underlying idea that it is not because things are visually challenging that the task cannot be performed = separation between QC for visual assessment and QC for algorithmic processing. Impact: Throw less data away

% So in total Artefact = N – I = Learnable component + Non Learnable residual = (R-N) + (N-I) = L + E

% sigma = uncertainty over reconstruction

% If the uncertainty over the reconstruction is correctly modelled, we want it to be high when the non learnable component is high so we want a strong correlation between sigma and E

\section{Related Work}

In \cite{Shaw2020}, the authors show that sources of uncertainty can be decoupled between task-specific uncertainty and uncertainties associated with different artefact subtypes, thus allowing the identification of the origin of image degradation. However, this approach only predicts uncertainty in the region of the segmentation prediction/label, not in the image overall. Therefore, to obtain a better estimate of overall image quality, we can pose the problem as an image reconstruction task, where the aim is to learn how to recover a clean image from a corrupted input. Combined with a probabilistic loss function, the uncertainty given the noisy data indicative of the non-recoverable component of the predicted image reconstruction informs the overall MR image quality. 

In the task of image reconstruction and denoising, Jiang et al. \cite{Jiang2017} use a CNN with residual L2 loss function on 2D slices to learn to remove Rician noise from MRI data. Duffy et al. \cite{Duffy2018} use the HighRes3DNet architecture combined with a discriminator loss to learn to remove motion artefacts from 3D MRI data, trained with simulated motion artefacts. Also using GANs, Ran et al. \cite{Ran2018}, learn to denoise MRI scans with  perceptual similarity losses to prevent the over-smoothing caused by mean-squared error (MSE) loss. Manjón et al. employ a two-stage approach: a 3D patch-based CNN for initial denoising, followed by a non-local averaging filter. In a similar approach to the one proposed in this work, Heinrich et al. \cite{Heinrich2018} use a U-Net architecture, although in 2D, with a residual connection between the input and the output to denoise CT scans. 

Additionally, the FastMRI challenge \cite{Knoll2020} explored MR reconstruction from under-sampled k-space, and the best-performing methods used complex and computationally expensive methods, such as deep network ensembles \cite{Hammernik2019} and iterative learning \cite{Liu2019}, with loss functions in both the k-space and the image domain. In this work the focus is shifted from the idea of performance -- the Bayesian loss function used is unlikely to provide state-of-the-art reconstructions -- to the question of self-awareness to errors as represented through uncertainty modelling. %NOT SURE IF WE SHOULD INCLUDE THE FOLLOWING... Even the best performing models may stumble in some complex cases, and in those cases, having appropriate indication over the prediction quality is essential. Our work, aims to provide, through uncertainty predictions, an indication of image quality that could potentially lead to learning-based real-time quality control systems for clinicians and radiographers.

\section{Method}

% We define the following notation:
% \begin{itemize}
%     \item[] I = Clean input image
%     \item[] N = Noisy input endallimage
%     \item[] R = Reconstructed prediction
%     \item[] E = Non-removable error (R-I)
%     \item[] L = Learn-able correction (N-R)
% \end{itemize}

In Bayesian deep learning, there are two main types of uncertainty:  \textit{epistemic} which is uncertainty in the model, and \textit{aleatoric} which depends on noise or randomness in the data. Furthermore, aleatoric uncertainty can be divided into two categories: \textit{homoscedastic} which is the task-dependent uncertainty, and \textit{heteroscedastic} which depends on the input data and can be predicted as a model output. We adopt a heteroscedastic noise model to capture uncertainty in the data. Heteroscedastic models assume that observation noise $\sigma^2$ can vary with the input $\mathbf{x}$, allowing for variable noise levels across the observation space \cite{Kendall2017_2}.

\begin{figure}[t!]
\centering
\includegraphics[width=.85\textwidth]{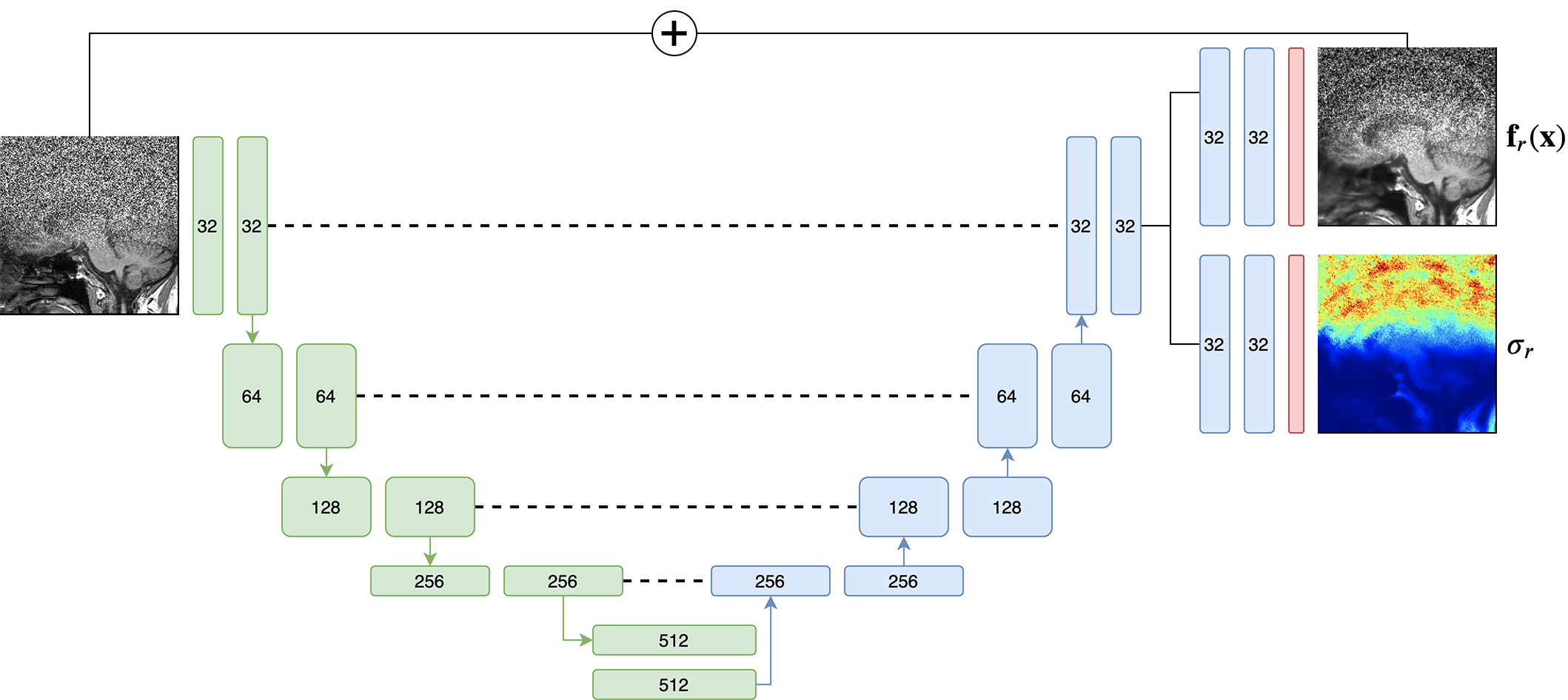}
\caption{3D U-Net with residual connection for image regression, outputting the reconstructed image $\mathbf{f}_r^{\mathbf{W}}(\mathbf{x})$ and predicted uncertainty $\sigma_r^2$.} \label{fig:unet}
\end{figure}

\subsubsection*{Network architecture and implementation details}
We use the updated 3D U-Net from \cite{Isensee2019} as the base architecture, as shown in Fig. \ref{fig:unet} implemented in NiftyNet \cite{NiftyNet}. The network is modified with two output heads, one for the prediction $\mathbf{f}_r^{\mathbf{W}}(\mathbf{x})$ and one for the uncertainty $\sigma_r^2$. In the task of image reconstruction, a residual connection is added between the input and output. 
All training was performed on a single Nvidia Tesla V100 GPU with Adam optimiser and batch size 1. Each model was trained with a learning rate of 0.0001 for 50k iterations.
\subsubsection*{Training data} was obtained from the Alzheimer's Disease Neuroimaging Initiative (ADNI) (\url{adni.loni.usc.edu}). Launched in 2003, ADNI attempts to assess whether medical imaging and biological markers and clinical assessment can be combined to measure the progression of Alzheimer's Disease. We use 272 MPRAGE scans that were deemed artefact-free, split into 80\% train, 10\% valid and 10\% test.

% \section{Experiments}

\section{Image Reconstruction Heteroscedastic Uncertainty}

% In this section we predict aleatoric heteroscedastic uncertainty for the task of image reconstruction given noisy input data.

\subsubsection*{Simulated artefacts}

We trained a model to regress clean images from simulated k-space artefacts \cite{Shaw2019} using the 3D U-Net architecture shown in Fig. \ref{fig:unet}. The data-dependent uncertainty was learnt using the probabilistic regression loss given by Eq. \ref{eq:regloss} \cite{Kendall2017}. 

\begin{equation}
    \mathcal{L}_{reg} =
    \frac{1}{2 \sigma_r^2} \left(
    \mathbf{y}_r - \mathbf{f}_r^{\mathbf{W}}(\mathbf{x})
    \right)^2
    + \frac{1}{2}
    \log \sigma_r^2
    \label{eq:regloss}
\end{equation}

\noindent where $\mathbf{y}_r$ is the ground-truth clean image, $\mathbf{f}_r^{\mathbf{W}}(\mathbf{x})$ is the output of the regression network with weights $\mathbf{W}$ and $\sigma^2_r$ is the predicted variance. Note, subscript r denotes the regression task variables. Eq. \ref{eq:regloss} derives from assuming the network predictions are normally distributed with mean $\mathbf{f}_r^{\mathbf{W}}(\mathbf{x})$ and variance $\sigma^2_r$ and computing log-likelihood results in this particular form.

\begin{figure}[!tbp]
\centering
\includegraphics[width=0.65\textwidth]{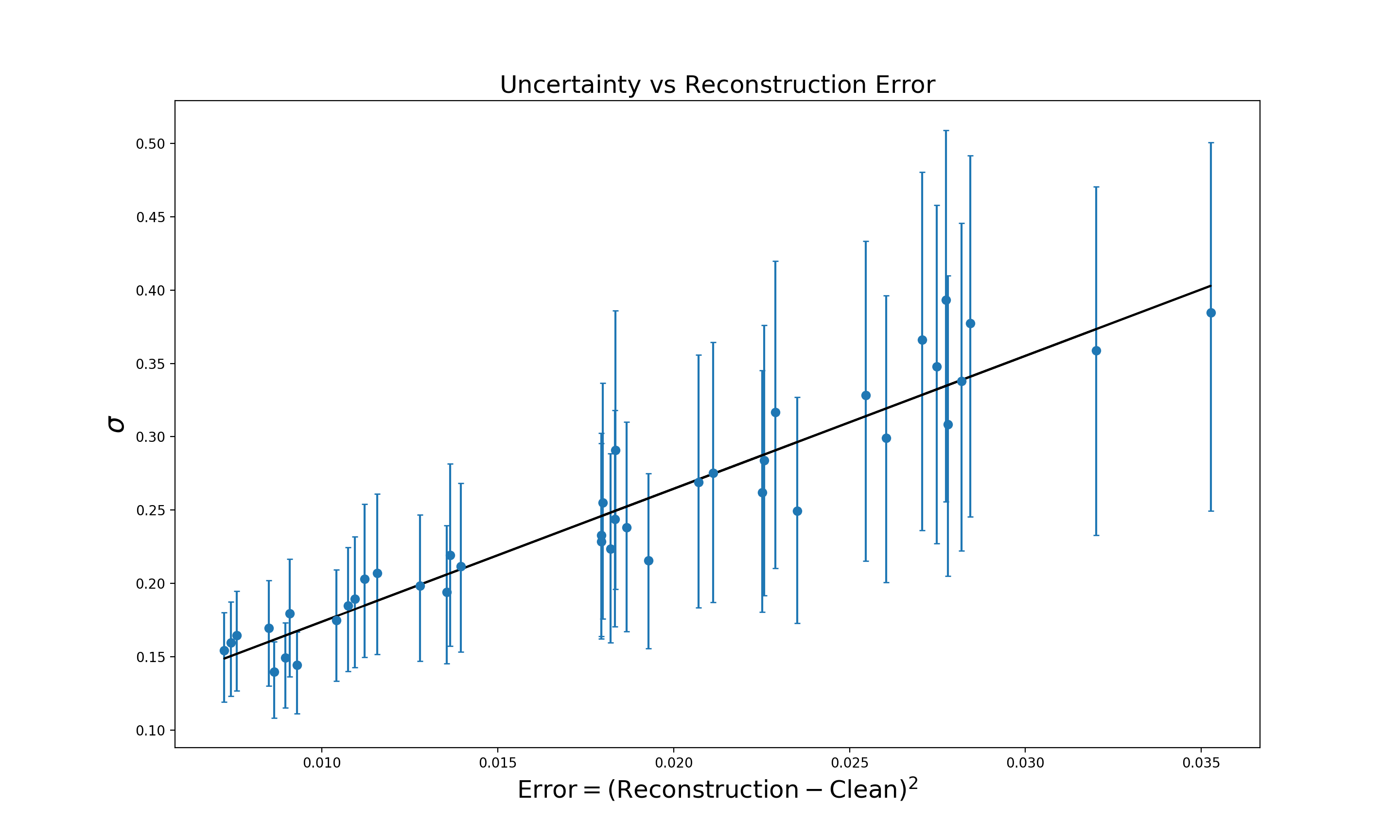}
\caption{Median image reconstruction error vs uncertainty as measured by the standard deviation $\sigma$. Errorbars represent the 1\textsuperscript{st} and 3\textsuperscript{rd} quartiles of the $\sigma$ distribution for a given image.} 
\label{fig:std_vs_E}
\end{figure}

For each image in the test set, 41 corrupted images with increasing levels of Rician noise were generated, spanning a signal-to-noise ratio from 10dB to -10dB (heavy noise). Inference was run on these images using the model trained with simulated artefacts. If the uncertainty over the reconstruction is correctly modelled, a strong positive correlation should be observed between the overall uncertainty $\sigma_r$ and the reconstruction error.
 Fig. \ref{fig:std_vs_E} presents for a given test case the association between the median of the predicted uncertainty $\sigma$ (1\textsuperscript{st} and 3\textsuperscript{rd} quartiles plotted as errorbars) against the non-removable image error $||\mathrm{reconstruction} - \mathrm{clean}||^2$ for increasing noise. $R^2$ in the presented case was 0.909.
 %0.929. Average $R2^$ over the 21 test cases was 0.844 and varied between between 0.686 and 0.936, confirming the strong alignment between uncertainty and non-recoverability. % Results when using the median 
 Average $R^2$ over the 21 test cases was 0.851 and varied between between 0.736 and 0.964, confirming the strong alignment between uncertainty and the non-recoverability of the image.

 % so we want a strong correlation between sigma and reconstruction error. 
.
\vspace{-20pt}

% Calculate the R2 per image and look at changes with level of noise

\subsubsection*{Real-world artefacts}

\begin{figure}[bp]
\centering
\includegraphics[width=0.5\textwidth]{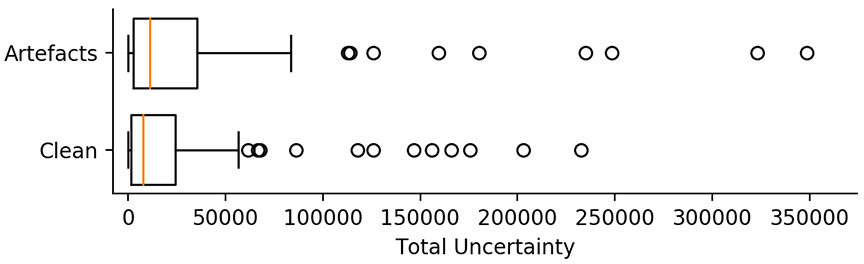}
\caption{Total uncertainty on real-world test-retest data from the heteroscedastic noise model trained with simulated k-space artefacts. The model predicts higher uncertainty on the artefact (test) set compared to the clean (re-test) set.} 
\label{fig:hbox1}
\end{figure}

To understand the benefit of training with simulated k-space artefacts when developing a QC model and the limitations encountered when limited by a specific set of observations, a similar model was trained using this time 106 real-world test-retest rigidly aligned pairs of T1-weighted images from the ADNI dataset. For each pair, one of the two scans had been labelled as ``containing artefacts."
%This should give us an idea of the upper bound of what is learnable from real-world data. 
While these artefacts are expected to be more realistic than synthetically generated ones, the amount and appearance variability of artefacts is limited by the size of the dataset. Furthermore, additional differences between paired images due to geometrical misalignment between the scans can potentially mask the artefacts' effects.  %We simply do not have access to a large corpus of real-world artefacts, and even if this could be acquired, it still would not have the scope of appearance variability that is possible with simulated k-space artefacts. %We can see how well we can still conserve that nice behaviour when training with synthetic data and testing in real-world cases.
When training only on real-world paired data, geometrical misalignment appeared to dominate the reconstruction uncertainty, resulting in localised regions of high uncertainty in specific misaligned areas while artefacted regions did not yield meaningful uncertainty measures. 

\begin{figure}[tbp]
\centering
\includegraphics[width=1.0\textwidth]{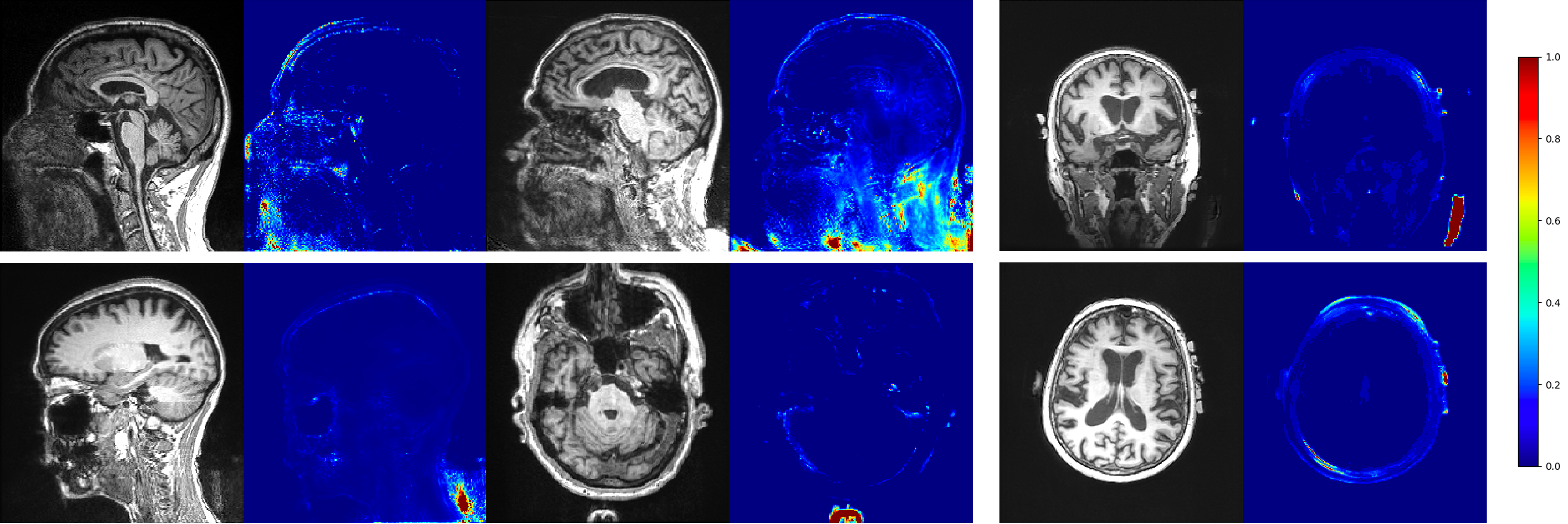}
\caption{Left: Reconstruction uncertainty predictions on real-world test-retest data from the model trained with simulated k-space artefacts. Right: Uncertainty predictions on images labelled as ``clean" by expert raters but still contain artefacts.} 
\label{fig:realworld}
\end{figure}

On the other hand, for the model trained with synthetic artefacts, perfect alignment between the artefact and clean images enables the uncertainty network to correctly model the noise in the data. A selection of these qualitative results is shown in Fig. \ref{fig:realworld} (Left). Note that high uncertainty is generally predicted in the artefacted regions. The total predicted uncertainty over test and re-test dataset is plotted in Fig \ref{fig:hbox1}, and we note that the model predicts higher uncertainty overall for the artefacted data. However, we notice that even the ``clean" dataset has a significant amount of total predictive uncertainty. This is because some individual images labelled as clean resulted in high uncertainty as they, in fact, had artefacts in them which had been missed by the human rater, as exemplified by Fig \ref{fig:realworld} (Right).

%Therefore, uncertainty due to noise and image artefacts is not adequately captured by the model, returning practically zero uncertainty in clearly artefacted regions. This highlights the difficulty of training with real-world artefact images. Whereas for the model trained with synthetic artefacts, we have perfect alignment between artefacted and clean data, therefore the uncertainty model is able to correctly reflect the noise in the data. 

\section{Multi-task Uncertainty Network}

\begin{figure}[tbh!]
\centering
\includegraphics[width=0.85\textwidth]{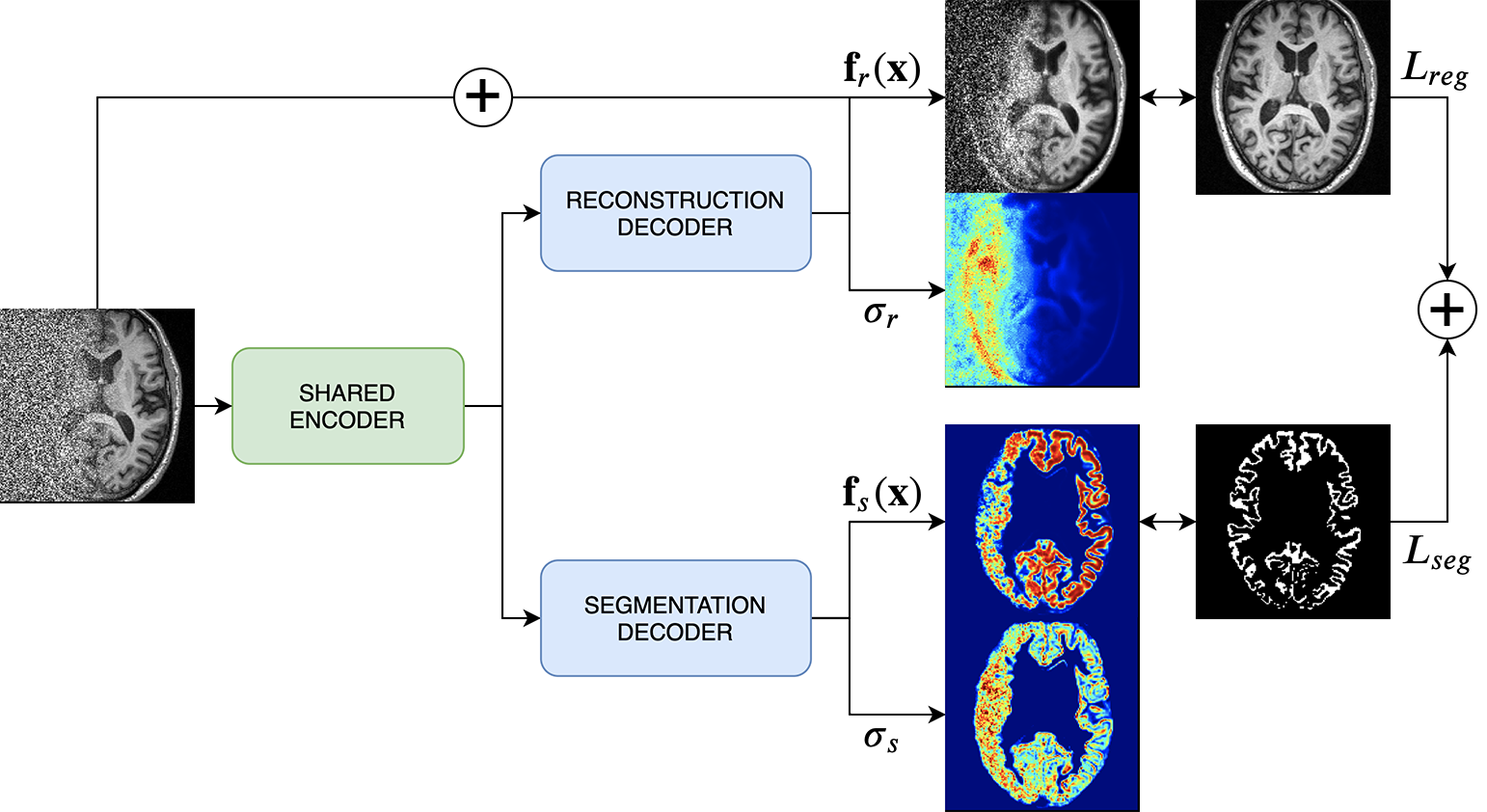}
\caption{Multi-task learning architecture. The predicted mean and variance are estimated for the regression and segmentation tasks, which share the encoder but have separate decoders. The task-specific loss functions $L_{reg}$ and $L_{seg}$ are combined to yield the total multi-task loss. The reconstruction uncertainty $\sigma_r$ captures the overall ``algorithmic quality'' of the input $\mathbf{x}$, while the segmentation uncertainty $\sigma_s$ reflects the specific ``task uncertainty."} 
\label{fig:multitask}
\end{figure}

In this experiment we extend our network from the single image reconstruction task to a multi-task setting; specifically image regression and cortical gray matter (CGM) segmentation. The network architecture is shown in Fig. \ref{fig:multitask}. The two tasks share a common 3D U-Net encoder but have separate decoders. The idea is that overall ``visual quality" is reflected by the image regression task uncertainty $\sigma_r$, while the ``algorithmic quality'' is reflected by the segmentation task uncertainty $\sigma_s$. Note, subscripts r and s denote the regression and segmentation task variables respectively.   

% See how well sigmaR and sigmaS relate with each other and total level of artefact to see if sigmaR good for generic QC perspective

For the segmentation task we use the weighted cross entropy loss as given by Eq. \ref{eq:segloss} \cite{Kendall2017}.
\begin{equation}
    \mathcal{L}_{seg} =
    \frac{1}{\sigma_s^2 }\mathrm{CE}\left(
    \mathbf{y}_s,\mathbf{f}_s^{\mathbf{W}}(\mathbf{x})
    \right)
    + \frac{1}{2}
    \log \sigma_s^2
    \label{eq:segloss}
\end{equation}

\noindent where $\mathbf{y}_s$ is the ground-truth CGM segmentation label, $\mathbf{f}_s^{\mathbf{W}}(\mathbf{x})$ is the output of the segmentation network with weights $\mathbf{W}$ and variance $\sigma^2_s$. The total multi-task loss is therefore the sum of the regression and segmentation losses:
\begin{equation}
    \mathcal{L}_{total} = \mathcal{L}_{reg} + \mathcal{L}_{seg}
\end{equation}

\begin{figure}[tbh!]
\centering
\includegraphics[width=1.0\textwidth]{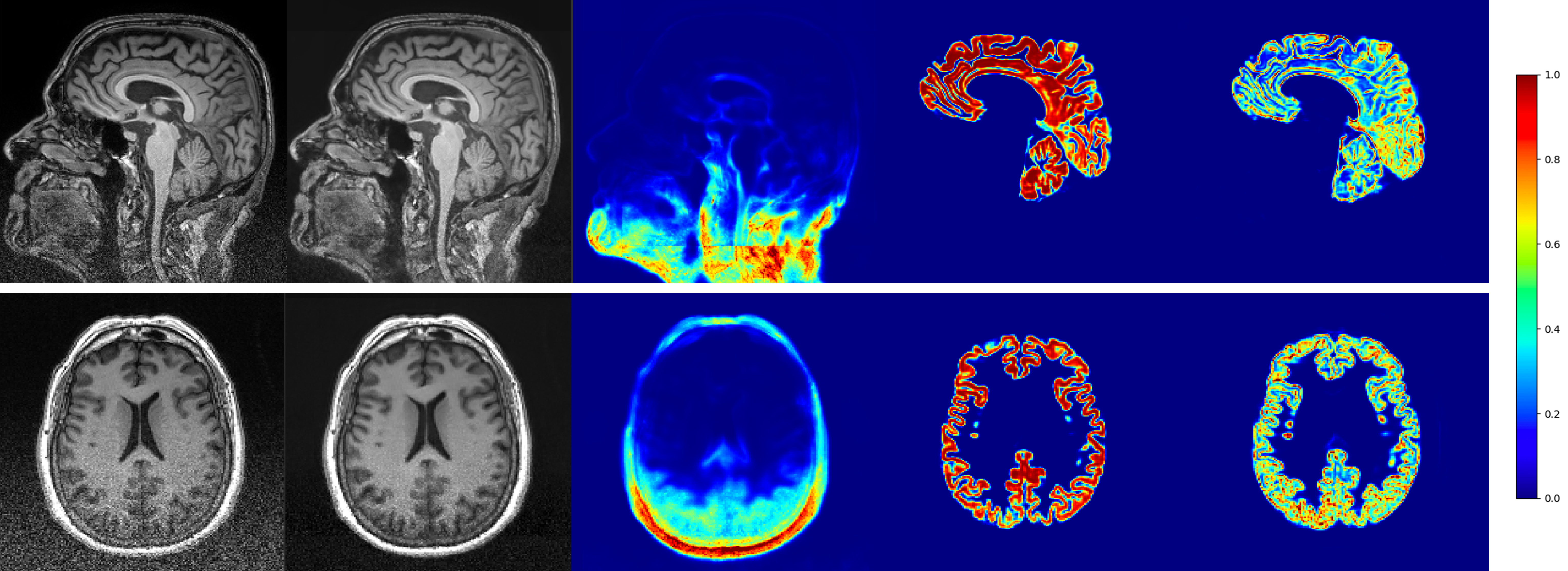}
\caption{Multi-task network predictions given partial image noise. From left to right: artefact input, recovered image, reconstruction uncertainty, CGM segmentation, and segmentation uncertainty. Note that the segmentation is largely unaffected, while the reconstruction uncertainty highlights the artefacted region.} 
\label{fig:preds}
\end{figure}

\begin{figure}[th!]
\centering
\includegraphics[width=0.5\textwidth]{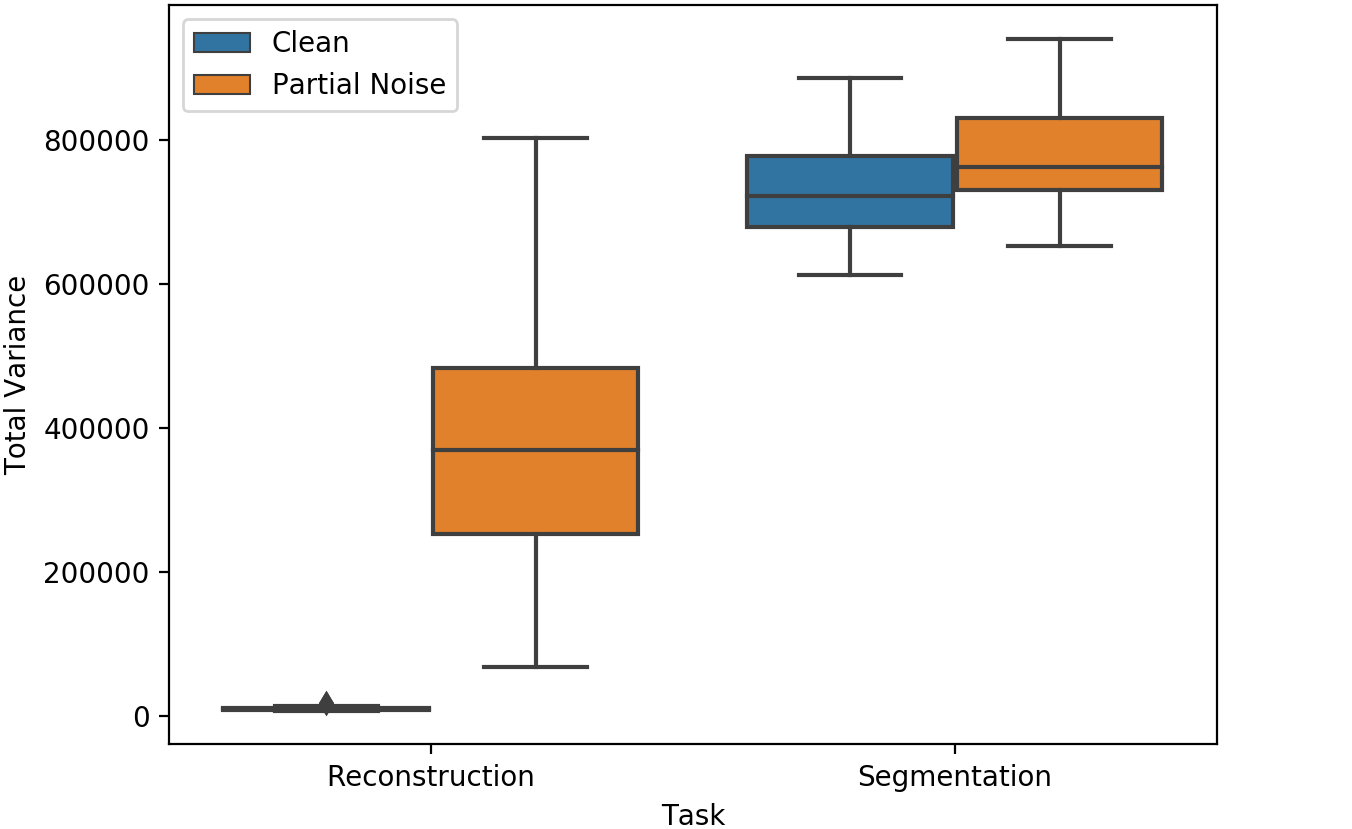}
\caption{Total uncertainty for the multi-task network given clean images and images with partial noise. The reconstruction uncertainty increase reflects the drop in perceptual image quality, while the segmentation task variance is similar.} 
\label{fig:box2}
\end{figure}

The multi-task network was trained with simulated k-space artefacts and we performed inference on the hold-out test set. To illustrate the difference between the two uncertainty predictions $\sigma_r$ and $\sigma_s$, we applied partial noise (rather than global image noise) to the test images so that the visual quality would degrade, yet the task of CGM segmentation would largely be, at least in-part, unaffected by noise. Examples of our predictions are shown in Fig. \ref{fig:preds}. We observe that reconstruction uncertainty reflecting the overall algorithmic quality increases on the artefacted data while the segmentation uncertainty reflecting the specific task uncertainty remains roughly constant, spiking only in areas of strongest overall algorithmic uncertainty associated to the task. This is shown quantitatively for the test set in Fig. \ref{fig:box2}. Thus, the overall visual quality can be described by the combination of the reconstruction uncertainty and the recoverable part of the corrupted image, while the algorithmic uncertainty is reflected by the segmentation task uncertainty.
\vspace{-10pt}
\vspace{-30pt}
\vspace{-30pt}

% \section{Results}

% Interpretation plot

%                 Limiting waste of scan plot – Assuming good presentation of sigma

%                 Plotting total artefact (N-I) vs ratio between learnable component and uncertainty (L/sigma)

% V stands for visual understanding – P for algorithmic processing

% 4 situations: sigma high TA high -> V- A- (very high artefact level with very high uncertainty)

%                          Sigma low TA high -> V- A+ (very high artefact level but most of it can be learnt and so no problem for processing)

%                          Sigma high TA low -> V+ A- (low artefact zone but not really learnable which makes the algorithmic processing difficult)

%                       Sigma low TA low -> V+ A+ (low artefact zone with certain prediction)

% \section{Discussion}
% This is a discussion of the results.

\section{Conclusion}

This work has explored a method of estimating MRI scan quality using heteroscedastic uncertainty in the task of image reconstruction. We have explored the relationship between visual quality assessment, i.e. what a human deems acceptable image quality, and algorithmic quality assessment through the associated predictive uncertainty. We propose that one should train a probabilistic image reconstruction network as an auxiliary task in order to decouple visual and algorithmic quality, hence enabling a real-time quality warning system for clinicians and radiographers.

%
% the environments 'definition', 'lemma', 'proposition', 'corollary',
% 'remark', and 'example' are defined in the LLNCS documentclass as well.
%
% For citations of references, we prefer the use of square brackets
% and consecutive numbers. Citations using labels or the author/year
% convention are also acceptable. The following bibliography provides
% a sample reference list with entries for journal
% articles~\cite{ref_article1}, an LNCS chapter~\cite{ref_lncs1}, a
% book~\cite{ref_book1}, proceedings without editors~\cite{ref_proc1},
% and a homepage~\cite{ref_url1}. Multiple citations are grouped
% \cite{ref_article1,ref_lncs1,ref_book1},
% \cite{ref_article1,ref_book1,ref_proc1,ref_url1}.
%\bibliographystyle{splncs04}


 %Author, F., Author, S.: Title of a proceedings paper. In: Editor,
 %F., Editor, S. (eds.) CONFERENCE 2016, LNCS, %vol. 9999, pp. 1--13.
 %Springer, Heidelberg (2016). %\doi{10.10007/1234567890}



% LNCS Homepage, \url{http://www.springer.com/lncs}. Last accessed 4
% \end{thebibliography}
\end{document}